\documentclass[twocolumn,showpacs,amssymb,prb]{revtex4}

\usepackage{graphicx}
\usepackage{dcolumn}
\usepackage{bm}

\newcommand*{\nhc}{(NH$_3$)$_{0.75}$NaK$_2$C$_{60}$}

\begin{document}

\title{$\mu$SR and SQUID Investigation of Superconductivity in (NH$_3$)$_{0.75}$NaK$_2$C$_{60}$}

\author{M.~Ricc\`o}%
\email{Mauro.Ricco@fis.unipr.it}
\homepage{http://www.fis.unipr.it/~ricco/}
\author{T.~Shiroka}%
\author{E.~Zannoni}%
\author{F.~Barbieri}%
\author{C.~Bucci}%
\affiliation{Dipartimento di Fisica and Istituto Nazionale di Fisica della Materia,\\
Universit\`a di Parma, Parco Area delle Scienze 7/a, 43100 Parma, Italy}
\author{F.~Bolzoni}%
\affiliation{Istituto Maspec-CNR, Parco Area delle Scienze, Loc.\ Fontanini, 43010 Parma, Italy}

\date{\today}%

\begin{abstract}
The family of superconducting fullerides (NH$_3$)$_x$NaK$_2$C$_{60}$ shows an 
anomalous correlation between $T_c$ and lattice parameter. To better understand 
the origin of this anomaly we have studied a representative $x=0.75$ compound
using SQUID magnetometry and $\mu$SR spectroscopy.

The lower critical field $H_{c1}$, measured by the trapped magnetization method, 
is less than 1 G, a very small value as compared with that of other fullerides. 
Muon spin depolarization in the superconducting phase shows also quite small local 
field inhomogeneities, of the order of those arising from nuclear dipolar fields. 
On the other hand, the 40 T value for $H_{c2}$, as extracted from magnetometry data, 
is comparable to that of other fullerides.
We show that these observations cannot be rationalized within the framework of the 
Ginzburg-Landau theory of superconductivity. Instead, the anomalous magnetic 
properties could be interpreted taking into account the role played by polaronic 
instabilities in this material.
\end{abstract}

\maketitle

\section{Introduction}
\label{sec:intro}
In fullerene based superconductors ammonia intercalates as a neutral molecule, 
without interacting with the host electronic system and preserving the 
superconducting properties of the material. 
Since it acts simply as a molecular spacer, a change in lattice 
parameters and an increase of unit cell volume are often observed\cite{Zhou93}.
As a consequence, the $t_{1u}$ conduction band of the
fullerene compound narrows and its density of states at the Fermi
level increases, thus determining an increment in the superconducting
transition temperature. An example of this mechanism has been reported
for Na$_2$CsC$_{60}$\cite{Zhou93} which gives 
(NH$_3$)$_4$Na$_2$CsC$_{60}$ after ammoniation, with a conspicuous 
increase in transition temperature from 10.5 to 29.6 K.

The intercalation of ammonia can also induce a metal-to-insulator 
transition as e.g.\ in NH$_3$K$_3$C$_{60}$,\cite{Iwasa96,Kitano02} 
where the superconductivity of the ammoniated compound can be restored only 
after the application of external pressure.\cite{Zhou95}

We will deal here with a family of fullerides such as 
(NH$_3$)$_x$NaK$_2$C$_{60}$ and (NH$_3$)$_x$NaRb$_2$C$_{60}$, whose 
precursors NaK$_2$C$_{60}$ and NaRb$_2$C$_{60}$ cannot exist as a single 
phase in normal conditions, but become stable only as ammoniated compounds.\cite{Shimoda96}
X-ray diffraction in these systems shows that  the NH$_3$-Na 
groups occupy the large octahedral sites\cite{Shimoda96} with a consequent 
off-centering of the Na$^+$ ions. Since their discovery, these compounds 
have revealed puzzling features concerning the relation between the 
superconducting transition temperature both with the lattice parameter 
as well as with the density of states at the Fermi level.
Indeed, the progressive removal of NH$_3$, accomplished by pumping on 
the sample above room temperature, results in a \textit{decrease} of the 
lattice parameters accompanied by an \textit{increase} of the 
superconducting transition temperature, a trend opposite to that observed in 
(NH$_3$)$_4$Na$_2$CsC$_{60}$.
In addition, we have recently shown\cite{Ricco01} that in a series of 
(NH$_3$)$_x$NaK$_2$C$_{60}$ compounds (with $0.5<x<0.8$) the Pauli-Landau spin 
susceptibility 
yields a \textit{lower} density of states at the Fermi level in compounds 
having a \textit{higher} $T_c$, in contrast with BCS or Migdal-Eliashberg 
predictions. 

It is therefore interesting to ask whether these anomalies 
are related to essential differences in the nature of superconductivity 
in \nhc\ with respect to most common superconducting fullerides. 
A possible non-conventional nature of the superconductivity in 
this system would manifest itself in the values of the critical parameters 
of its superconducting phase: the London penetration depth $\lambda$ and the 
lower and upper critical fields, $H_{c1}$ and $H_{c2}$ respectively. Such fundamental 
parameters have never so far been measured and we will tackle the problem by 
providing their values and discussing the implications. 

From the available data in the literature, the other so-called ``normal''
fullerene based superconductors appear to be extreme type-II 
superconductors\cite{Holczer91,Sparn92} characterized by 
$\kappa = \lambda/\xi \gg 1$, where $\kappa$ is the Ginzburg-Landau parameter. 
Our results will be first analyzed in this framework and, in case of discrepancies,
alternative suggestions will be offered.

When $\lambda \gg \xi$, both of these fundamental lengths can be easily 
extracted from measurements of the lower ($H_{c1}$) and the upper ($H_{c2}$) 
critical magnetic fields or, more precisely, from their extrapolated values 
at zero temperature. Roughly speaking, $H_{c2}$, the field at which the 
transition from the superconducting to the normal state occurs, corresponds 
to the field at which one quantum of magnetic flux 
$\Phi_0=hc/2e \simeq 2 \times 10^{-7}$ G cm$^2$ extends 
over the coherence area of an electron pair, so that:
\begin{equation}
\label{eq:hc2}
H_{c2}(0) = \frac{\Phi_0}{2\pi\xi^2} ,
\end{equation}
from which the Ginzburg-Landau coherence length $\xi$ can be determined. 
On the other hand, the knowledge of $H_{c1}$, the field at which the 
magnetic flux starts to penetrate the sample, allows the computation of 
the penetration depth $\lambda$ by using the well known equation 
(valid for $\kappa \gg 1$):
\begin{equation}
\label{eq:hc1}
H_{c1}(0) = \frac{\Phi_0}{4\pi\lambda^2}\ln\kappa .
\end{equation}

In addition to independent measurements of $H_{c1}$ and $H_{c2}$, as obtained
by standard SQUID magnetometry, we will provide also the value for $\lambda$,
which yields a stringent check of the validity of eq.~(\ref{eq:hc2}) and
(\ref{eq:hc1}).
Since $\lambda$ represents the transverse extension of the vortices in the 
Abrikosov intermediate phase, its value can be also determined from the local 
magnetic field distribution. 
It is well known that Muon Spin Rotation ($\mu$SR) gives reasonable estimates
of $\lambda$, even for irregular flux line lattices. Details on SQUID and 
$\mu$SR measurements are given in the next experimental section.

\section{Experimental}
\label{sec:experiment}
The samples were prepared following the procedures outlined 
in Ref.~\onlinecite{Shimoda96}.
Stoichiometric amounts of alkali metals
(Aldrich, 99.95\%) and C$_{60}$ (Southern Chem.\ 99.5\%) were dissolved 
in anhydrous ammonia (Aldrich, 99.99+\%) at 230 K. After the reaction had
taken place, the temperature was slowly increased until the ammonia
was completely evaporated. The successive pumping at 120$^\circ$C for 
30 minutes yielded the compound \nhc; the sample was then
annealed at 100$^\circ$C for 10 days. Its transition temperature was 
$T_c = 12$ K and it showed a 20\% superconducting fraction, indicative 
of bulk superconductivity. The ammonia concentration $x=0.75$ was 
determined from the superconducting transition temperature by interpolating 
the $T_c-x$ data reported in Ref.~\onlinecite{Shimoda96}.

Evaluation of the granulometry of the samples was performed using 
a scanning electron microscope (SEM), whose micrographs indicate 
an average size of the particles $d \sim 2$ $\mu$m.

DC magnetometry measurements were performed with a Quantum Design
SQUID magnetometer equipped with a home built Helmholtz cube which
surrounded the whole instrument body, thus allowing a reduction of the
residual field on the sample to less than 2 mG. The SQUID
superconducting magnet was cooled from RT to liquid He temperature in
zero external field.
For a good thermal contact even at low temperatures the sample was
sealed in a long quartz tube under 1 mbar of He atmosphere. The sample
was suspended in the middle of the tube, whose length was so chosen as 
to have always a tube portion face the SQUID coils, even when the sample 
had to move in and out of them during the magnetic moment measurement.
This expedient allowed an accurate subtraction of the quartz diamagnetic 
contribution.

$\mu$SR, which measures the spin precession of
implanted muons, is very sensitive to local magnetic fields and
therefore it constitutes a valuable technique for our purposes. Indeed,
when the applied transverse field exceeds $H_{c1}$, the field distribution
of the flux line lattice will damp the muon spin precession signal, hence 
the penetration depth can be readily extracted\cite{MacFarlane98}
from the damping rate.

In common metals and superconductors all the implanted muons
usually sit interstitially and, being screened by conduction electrons, 
they will not form any paramagnetic bound state (muonium). Hence, their
precession frequency in an external magnetic field remains that of a
free particle (diamagnetic muon). In C$_{60}$ based superconductors, besides
this major component there is an additional part of implanted muons 
(typically 10--20\%) which will form endohedral muonium (located inside 
a C$_{60}$ molecule), whose precession frequency is much 
higher than that of diamagnetic muon.
Here we are interested only in the majority of muons that come at rest in 
the \textit{fcc} lattice interstices (the precise location is not well known) 
and precess as diamagnetic muons, i.e.\ with a gyromagnetic ratio 
$\gamma_{\mu}=13.55$ kHz/G. 

When a superconductor is in the intermediate state, i.e.\ 
$\mu_0H_{c1}<B<\mu_0H_{c2}$, the muon precession signal will be damped 
by the inhomogeneous magnetic field distribution of the vortices. The
expected damping profile (or the corresponding lineshape in Fourier space) 
in the case of a triangular flux-line lattice has been
computed\cite{Brandt88} and, for single crystals, also successfully 
measured. Unlike single crystals, polycrystalline or powder
materials exhibit a smeared out magnetic field distribution, with the 
consequence that the $\mu$SR line will assume a Gaussian shape
($\sigma_{\text{sc}}\sim0.1$--0.6 $\mu$s$^{-1}$ for fullerides) below 
$T_c$.\cite{MacFarlane98} 
The established relation between the $\mu$SR damping rate 
$\sigma_{\text{sc}(0)}$ and the internal field rms deviation 
$\Delta B$ is given by $\Delta B = \sigma_{\text{sc}}(0)/2\pi\gamma_{\mu}$. 
Once $\Delta B$ is known from an experiment that measures 
$\sigma_{\text{sc}}(0)$ in the appropriate intermediate field range
$\mu_0H_{c1}<B<\mu_0H_{c2}$, the penetration depth $\lambda$ is given 
by:\cite{Brandt88}
\begin{equation}
\label{eq:lamdB}
\lambda \approx 3.71\times 10^{-3}\cdot%
\left[\frac{\Phi_0^2}{(\Delta B)^2}\right]^{1/4}.
\end{equation}

$\mu$SR experiments were performed on the EMU spectrometer of the ISIS
Facility (Rutherford Laboratory, UK). The pulsed nature of the muon beam
sets an upper frequency cut-off that prevents the direct observation 
of high muonium frequencies. Nevertheless, the interesting fraction of 
diamagnetic muons will precess well within the pass-band of the spectrometer 
and therefore could be readily measured.
In our case, the sample ($\sim 500$
mg) was pressed inside an air tight aluminum cell equipped with a
thin (75 $\mu$m) Kapton window. A pure silver foil was put both directly 
behind the sample and around the cell window to make it easy to subtract 
the signal coming from the sample holder.

\section{SQUID Measurements}
\label{sec:squid}
\subsection{Lower critical field}
\begin{figure}
\includegraphics[width=0.45\textwidth]{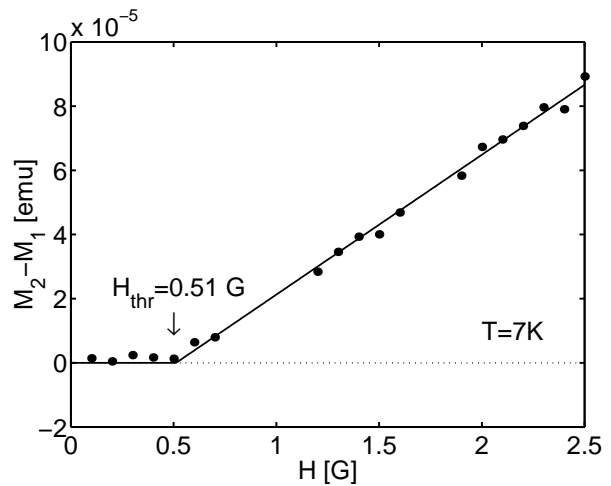}
\caption{\label{fig:hc1_trapped}The trapped magnetization $M_2-M_1$ as a 
function of the applied field $H$ in \nhc. The lower critical field $H_{c1}$ 
is determined  from the onset of the magnetization which starts is trapped 
at $H=H_{\text{thr}}$.}
\end{figure}

In spite of many previous measurements of the lower critical field by
SQUID magnetometry in fullerides,\cite{Sparn92,Politis92a,Politis92b,%
Holczer91,Buntar92,Buntar95a,Buntar95b,Buntar96,Buntar97,%
Buntar98} its precise determination is still controversial due to 
considerable experimental difficulties. 
The simplest way to measure $H_{c1}$ consists in observing the field at which
the $M=M(H)$ curve starts to deviate from linearity. Unfortunately data
on fullerides never show a good linearity and this brings to an
overestimate of the $H_{c1}$ value. Alternative methods based on Bean's
critical state model\cite{Bean62} give quite different 
values.\cite{Buntar95a} Recently it was shown\cite{Buntar96} 
that the measurement of the trapped magnetization in the intermediate
phase yields  a more reliable determination of $H_{c1}$. The procedure
is as follows:  first the sample is cooled in zero field from above 
$T_c$ and its initial magnetic moment ($M_1$) is measured (ideally it 
should be zero, but a residue always exists). Then a magnetic
field $H$ is applied for at least 30 s and, after switching it off, the
final moment ($M_2$) is measured. When the applied field exceeds
a threshold value $H_{\text{thr}}$ the magnetic flux is trapped 
inside the sample. This trapped magnetization, given by the difference 
$M_2-M_1$, is then plotted against the applied field $H$, as shown in 
Figure~\ref{fig:hc1_trapped} for measurements on \nhc\ performed at 7 K. 
The increase of the trapped magnetization above $H_{\text{thr}}$, which 
follows a linear behavior at all the investigated temperatures, allows 
a much more precise determination of $H_{c1}$ than alternative procedures. 
The fitted value for trapping onset 
(0.51 G for the linear fit shown in Figure~\ref{fig:hc1_trapped}) is related 
to the lower critical field value by $H_{c1} = H_{\text{thr}}/(1-n)$, where $n$ 
is the demagnetization factor.  Preliminary measurements on powdered samples 
gave essentially the same results, although the trapped magnetization values 
were rather scattered around the (same) fit line (see Figure~\ref{fig:hc1_trapped}).
The use of (weakly compressed) pellets remarkably reduced the spread 
without an appreciable change in $H_{\text{thr}}$, confirming that 
the demagnetization factor of a set of independent spheres ($n=1/3$), 
correctly adopted for powders,\cite{Buntar96} is appropriate also for pellets.

\begin{figure}
\includegraphics[width=0.45\textwidth]{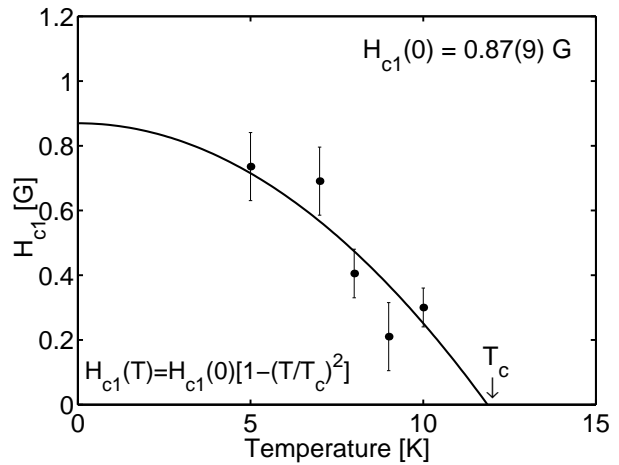}
\caption{\label{fig:hc1_T}Measured $H_{c1}$ values as a function of temperature. 
The extrapolation of the parabolic fit at $T=0$ yields $H_{c1}(0)=0.87(9)$ G.}
\end{figure}

The results of several $H_{c1}$ measurements at different temperatures are
illustrated in Figure~\ref{fig:hc1_T}. The zero temperature value 
$H_{c1}(0)=0.87$ G was then extrapolated from a parabolic fit (BCS weak 
coupling gave a similar result and experimental errors do not allow to 
distinguish between the different behaviors).

The striking feature about the $H_{c1}(0)$ is its low value with respect 
to that of  other fullerides, shown in Table~\ref{tab:fullerides} for 
a comparison. Even an unphysical linear fit of the data would at most 
yield an extrapolated value $H_{c1}(0)\sim 1.2$  which still is one order 
of magnitude smaller than  $H_{c1}(0)$ for the quoted fullerides.  The obvious 
suspicion that the powdered nature of the sample affects the measured values was 
addressed in a previous experiment, reported in Ref.~\onlinecite{Buntar96}, where 
a powdered RbCs$_2$C$_{60}$ sample and a single crystal were compared and found to 
have the same lower critical field value. Hence, the measured low $H_{c1}$ value 
given here will be considered an intrinsic feature of the compound. 
\begin{table}
\caption{\label{tab:fullerides}Comparison of lower critical fields 
$H_{c1}$ for several alkali metal doped fullerides (powders denoted by asterisk).}
\begin{ruledtabular}
\begin{tabular}{lcc}
Compound	&$H_{c1}(T=0)$ [G]	&Ref.\\
\hline
Rb$_3$C$_{60}$	&$\sim 50$	&\onlinecite{Buntar92}\\
Rb$_3$C$_{60}$	&13*		&\onlinecite{Buntar96}\\
RbCs$_2$C$_{60}$&$\sim 80$	&\onlinecite{Buntar95a}\\
RbCs$_2$C$_{60}$&16*		&\onlinecite{Buntar96}\\
K$_3$C$_{60}$	&12*		&\onlinecite{Buntar96}\\
\nhc\		&0.87(9)*	&this work\\
\end{tabular}
\end{ruledtabular}
\end{table}

\subsection{Upper critical field}
\begin{figure}
\includegraphics[width=0.45\textwidth]{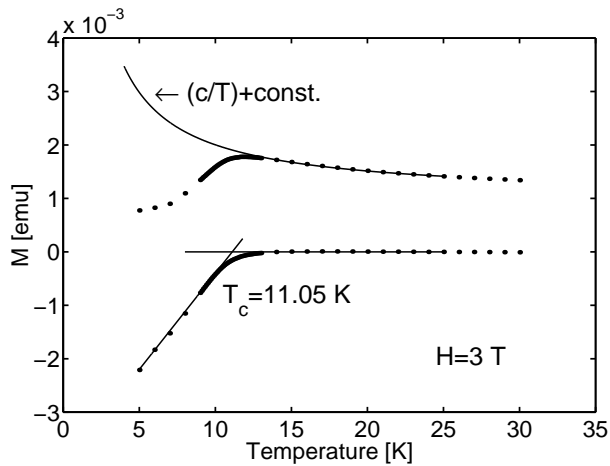}
\caption{\label{fig:measHc2}Upper graph: Field cooled magnetization of 
\nhc\ at $H=3$ T: the fitted  curve accounts for paramagnetic impurities and 
Pauli contributions. Lower graph: After the subtraction of the above
mentioned contributions, $T_c$, at $H_{\text{appl}}=H_{c2}$, is determined 
from the intersection of the two linear fits.}
\end{figure}

The upper critical field $H_{c2}$ can be determined from the temperature
dependence of the field cooled magnetization.\cite{Buntar95a} 
In this case the sample is cooled in an externally applied field $H$
(0.5 T $<H<$ 5.5 T) starting from $T>T_c$ and its magnetic moment is
measured. The value of the external field corresponds to $H_{c2}$ when 
the temperature equals the relative $T_c$. Figure~\ref{fig:measHc2} 
shows an example of such a measurement in a 3 T magnetic field. 

\begin{figure}
\includegraphics[width=0.45\textwidth]{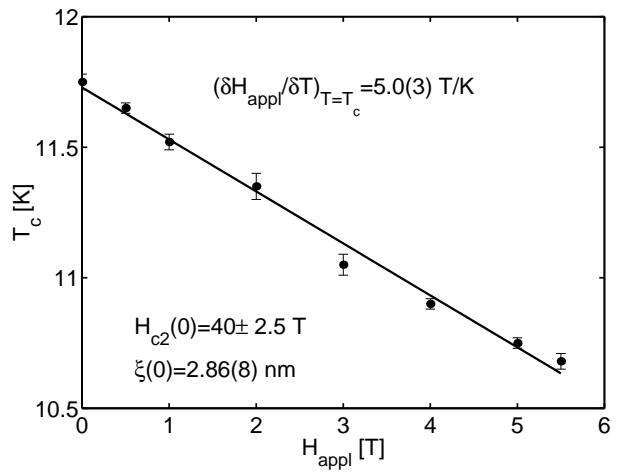}
\caption{\label{fig:TvsHc2}Critical temperature dependence on $H_{\text{appl}}$. 
The upper critical field $H_{c2}(0)=40 \pm 2.5$ T is found from the extrapolated 
slope $\partial H_{\text{appl}}/\partial T_c$ using formula~(\ref{eq:whh}).}
\end{figure}

The data above $T_c$ were fitted to a Curie behavior coming from paramagnetic impurities, 
to which one must add a temperature independent component resulting from Pauli, Landau 
and core contributions to susceptibility (upper part of the
figure).\cite{Ricco01} After subtraction of all these
contributions the curve shown in the lower part of Figure~\ref{fig:measHc2} 
is obtained. $T_c$ was estimated as the temperature at which the linear
interpolation of the data in the superconducting state intersects the
normal state baseline. The observed superconducting transition
temperature $T_c(H)$ decreases on increasing $H_{\text{appl}}$. In 
Figure~\ref{fig:TvsHc2} we report the dependence of the critical temperature on 
$H_{\text{appl}}$. 
Unfortunately, the maximum field available in our conventional 
SQUID magnetometer ($H_{\text{max}}= 5.5$ T) does not allow us 
to investigate the full range of $T_c=T_c(H_{\text{appl}})$ dependence so that we 
have to resort to extrapolation. The extrapolated field value at zero 
temperature (relevant for $\xi$ calculation) is usually extracted from the 
slope of the observed linear behavior using the Werthamer-Helfand-Hohenberg 
(WHH) formula:\cite{Werthamer66}
\begin{equation}
\label{eq:whh}
H_{c2}(0) = 0.69\;T_c\cdot\left.\frac{\partial H_{c2}}{\partial T}\right|_{T=T_c}.
\end{equation}

The value for the derivative is $5.0 \pm 0.3$ T/K and from eq.~(\ref{eq:whh}) we 
find $H_{c2}(0)=40\pm 2$ T. If we use eq.~\ref{eq:hc2}, we can extract a coherence 
length $\xi=2.86(8)$ nm. Both of these values, although different, do not appear to 
be inconsistent  with those of other C$_{60}$ based superconductors.\cite{Baenitz95} 

Similarly to the measurement of $H_{c1}$ it is important to examine possible 
factors that could affect the estimated value for $H_{c2}$:\\
a) According to analogous measurements on K$_3$C$_{60}$\cite{Boebinger92} the 
parabolic $H_{c2}(T)$ dependence predicted by the WHH theory\cite{Werthamer66} 
was not observed: the use of such theory to extract $H_{c2}(0)$ was shown to 
produce an underestimated value. By guessing from Ref.~\onlinecite{Boebinger92} 
a plausible enhancement factor of 2 in our datum, we point out that 
eq.~(\ref{eq:hc2}) would predict a still smaller $\xi$ ($\sim 2$ nm), which 
has an important role in the discussion that will follow.\\
b) It is known that sample granularity enhances the measured $H_{c2}(0)$ values 
due to the onset of 0-D fluctuations, as detected in conductivity measurements 
in K$_3$C$_{60}$.\cite{Hou94} This effect is expected to become dramatic when 
the grain size $d$ becomes comparable to $\xi$. Indeed, also conventional 
superconductors like Al, in a suitable fine granular form, can have an upper 
critical field nearly two orders of magnitude larger than that of bulk
samples.\cite{Abeles67} The granularity of the material we have investigated, 
however, involves an average particle size of $\sim 2$ $\mu$m, which is more 
than three orders of magnitude larger than any estimate for $\xi$, thus allowing
us to definitely exclude any appreciable $H_{c2}$ enhancement effect due
to granularity.

In conclusion we can state that the coherence length we find from $H_{c2}$ 
measurements is accurate and, due to its inverse square-root dependence 
from $H_{c2}$, even large uncertainties in the determination of the latter 
would not appreciably affect the $\xi$ value.

\section{Muon Spin Rotation Measurements}
\label{sec:musr}
As briefly described in Sec.~\ref{sec:experiment}, $\mu$SR damping rate 
measurements yield a reasonable value for the penetration depth $\lambda$, 
even for an irregular flux-line configuration, typical for powders or 
polycrystalline samples.
In our experiment the sample was field-cooled in an external transverse
field of 50 G from above $T_c$ and the muon precession histograms measured 
at fixed $T$. 
The presence of two precessing signals at all  temperatures indicated 
that a fraction of the total muons came at rest in the sample holder 
while the other in the sample. The two components could be easily singled 
out thanks to their differences not only in amplitude, but mainly in their 
precession frequency (diamagnetic shift due to superconducting material) 
and damping rate (internal field second moment due to flux-line lattice).
In accounting for the whole expected signal, a missing fraction was detected: 
it is well known that this fraction is entirely due to the formation of 
endohedral muonium --- i.e.\ muonium atoms at rest within the C$_{60}$ cage --- 
while the formation of muonium adduct radicals is inhibited in C$_{60}^{n-}$ 
compounds. This fraction is ``missing'' because, at the applied field of 50 G, 
its characteristic frequency is too  high  to be observable at the ISIS facility. 
In passing, it must be pointed out that this missing muonium fraction is not 
interesting for the present experiment, where we look for the information 
yielded by the unbound muons in the material's interstitial sites.

\begin{figure}
\includegraphics[width=0.45\textwidth]{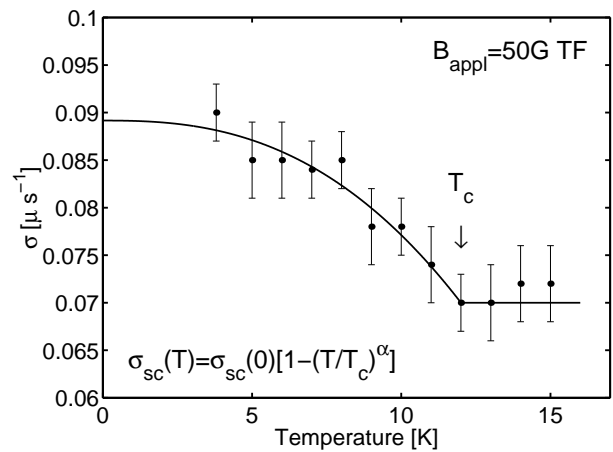}
\caption{\label{fig:sigmaT}$\mu$SR Gaussian decay rate as a function of 
temperature in a 50 G transverse field. The fit of data below $T_c$ yields 
$\sigma_{\text{sc}}$, from which the field distribution inside the sample is found. 
The penetration depth $\lambda$ is then calculated using eq.~(\ref{eq:lamdB}).}
\end{figure}

The signal due to the sample shows a Gaussian decay with a decay rate  
$\sigma$, which was fitted using the function $\sin(\omega t+\phi)\exp(-\sigma^2t^2/2)$. 
Figure~\ref{fig:sigmaT} shows the fitted values for $\sigma$ in the temperature 
range 4 $<T<$ 15 K. The temperature independent residual value as observed above 
$T_c$ is due to the magnetic field distribution of the randomly oriented 
nuclear dipoles ($^1$H, $^{14}$N, $^{23}$Na, $^{39,40,41}$K, $^{13}$C). 

The additional broadening below $T_c$, due to the flux-line lattice formation,
is smaller than the  nuclear term and much smaller than  that found
in similar fullerides such as Rb$_3$C$_{60}$ and
K$_3$C$_{60}$.\cite{MacFarlane98} The superconducting contribution and
the normal state nuclear dipole broadening add in quadrature, hence the
former ($\sigma_{\text{sc}}$) can be extracted. The temperature
dependence of $\sigma_{\text{sc}}$ was fitted to the phenomenological
temperature dependence:
$\sigma_{\text{sc}}(T)=\sigma_{\text{sc}}(0)[1-(T/T_c)^{\alpha}]$,
which yields $\sigma_{\text{sc}}(0)= 5.5\cdot10^{-2}$ $\mu s^{-1}$ and
$\alpha=2.55$. The same experiment was repeated applying a 100 G
transverse field; in this case no significant increase in
$\sigma_{\text{sc}}$ was observed.  The value for the London
penetration depth $\lambda$ can be extracted from eq.~(\ref{eq:lamdB}) 
which  gives $\lambda=1.40$ $\mu$m. This value is more than two times 
larger than that of other superconducting fullerides. In thinking of 
effects that could enhance the measured $\lambda$ a possible candidate 
is the granulometry of the sample, when the particles size is not much 
larger than  $\lambda$. But this is just our case, where being 
$d \sim \lambda$, we do expect strong surface effects which tend to 
increase the measured $\lambda$ value with respect to that of a bulk
sample. Although we have no means to implement a quantitative correction 
to the measured value, we can definitely state that it represents an upper 
limit for the real penetration depth, which certainly cannot be 
larger than 1.4 $\mu$m.

\section{Discussion}
\label{sec:discuss}
In the previous section we described the independent measurements of
$H_{c1}$, $H_{c2}$, and $\lambda$ and anticipated that they would be
used within the framework of the GL-Abrikosov equations~(\ref{eq:hc2}) and 
(\ref{eq:hc1}). In principle, the knowledge of $H_{c1}$ and $H_{c2}$ values 
is sufficient to extract the coherence length $\xi$ and the penetration 
depth $\lambda$.  
The additional experimental value of $\lambda$ from $\mu$SR measurements
helps in checking  the internal consistency. From eq.~(\ref{eq:hc2}) the
experimental value of $H_{c2} = 40$ T yields $\xi = 2.86(8)$ nm; with
this  value and the experimental result for $H_{c1}= 0.87$ G 
eq.~(\ref{eq:hc1}) yields $\lambda = 3.82$ $\mu$m. 
If only $H_{c1}$ and $H_{c2}$ were measured, the values for $\xi$ and 
$\lambda$ (although the latter seems considerably larger than the value 
found in other superconducting fullerides) would appear acceptable for 
an extreme type-II superconductor. 

The picture appears inconsistent, 
however, since we must account for the \textit{measured} value of $\lambda = 1.4$ 
$\mu$m, which is nearly three times smaller than that predicted by 
eq.~(\ref{eq:hc2}) and (\ref{eq:hc1}). 
Such a discrepancy is by no means trivial because, in order to weaken it, 
one has to make unphysical assumptions on the other measured parameters, 
i.e.\ the two critical fields. Indeed, as can be seen from eq.~(\ref{eq:hc2}) 
and (\ref{eq:hc1}), even quite a small variation in $\lambda$ will determine
orders of magnitude variations in the critical fields.

As an additional check for the observed discrepancy we have also used
some approximate analytical solutions of the GL equations. 
Specifically, we compute the muon spin depolarization rate 
$\sigma_{\text{sc}}$, which can easily be compared with the experimental 
value. The calculation requires a numerical solution of the GL equations 
which was performed using the efficient algorithm developed in 
Ref.~\onlinecite{Brandt97}. This fast iterative variational procedure, 
which minimizes the free energy density $f$, gives as an output the 
local magnetic field distribution $B=B(x,y)$. The second moment of 
the local fields, $\Delta B$, yields the damping rate ($\sigma_{\text{sc}}$)
of the $\mu$SR.
\begin{figure}
\includegraphics[width=0.45\textwidth]{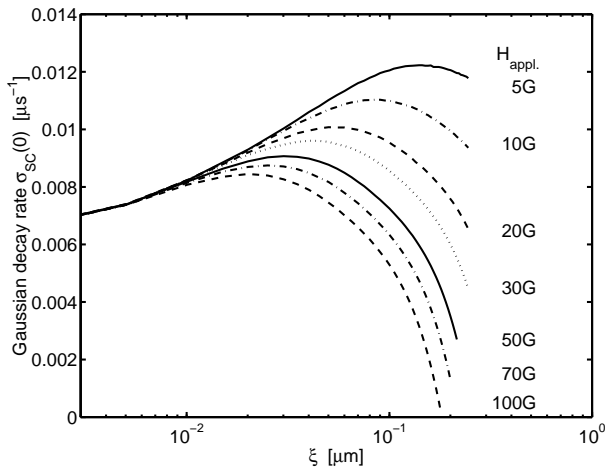}
\caption{\label{fig:sigma_vs_xi}$\mu$SR signal decay rate
$\sigma_{\text{sc}}$  as a function of coherence length $\xi$ as
obtained from the numerical solution  of the GL equations. The
different curves correspond to several values of applied  magnetic
field; the lower critical field value was fixed at $H_{c1}=0.87$ G.}
\end{figure}

The results of this calculation are shown in Fig.~\ref{fig:sigma_vs_xi} 
which reports the dependence of $\sigma_{\text{sc}}$ on the coherence 
length $\xi$ as predicted by the GL equations for a fixed $H_{c1}=0.87$ G. 
The different curves refer to different values of the applied field in 
the range 5 G $< B < 100$
G. It appears evident that the maximum predicted value for the muon
decay rate at the applied field of 50 G is $9\cdot10^{-3}$
$\mu$s$^{-1}$, much smaller than the measured value of
$55\cdot10^{-3}$ $\mu$s$^{-1}$ and, in any case, too small to be
measured. This confirms that the $H_{c1}$, $H_{c2}$ and $\lambda$
values measured in \nhc\ cannot be representative of a superconducting
system described by the Ginzburg-Landau theory. In addition, the fact
that a series of samples with different ammonia content have critical
temperatures inversely proportional to the density of states at the
Fermi level, as recalled in the Introduction, is consistent with this
conclusion.

Even though the phonon-mediated superconductivity in fullerides is 
well established, we suggest that  polaron-based theories of
superconductivity can correctly describe some of these systems, 
such as those considered in this work. Unlike the BCS or
Migdal-Eliashberg alternatives, the polaronic approach  does not 
require the phonon energy scale to be much smaller than that of 
electrons. In fullerides, indeed, typical phonon energies involved in
the superconducting coupling are or the order of 0.15 eV, due to the 
intra-molecular $H_{\text{g}}$ modes of C$_{60}$.\cite{Gunnarsson97}  
According to the early predictions of N.L.\ Bulaevskii and
co-workers\cite{Bulaevskii84} for the superconducting properties of
systems with local pairs, critical fields and critical length values
are expected to be significantly different from those of ordinary BCS 
superconductors; in particular, the lower critical field $H_{c1}$ is 
expected to be much smaller and the penetration depth $\lambda$ much
larger than the respective BCS counterparts. 
More recently, a polaron-based analysis has been  formulated,
with the inclusion of  non adiabatic effects in the Migdal-Eliashberg
theory, in order to better predict transition temperatures and
photoelectron spectra of fullerides.\cite{Alexandrov96}  The presence
of charge instabilities or fluctuations (charge disproportion, charge
density waves etc.)  predicted in polaronic systems manifests itself,
indirectly, in superconducting fullerides in the following
circumstances: a) the  NMR detection of a spin gap in the
Na$_2$CsC$_{60}$ superconductor as due to  the presence of Jahn-Teller
distorted C$_{60}^{(2,4)-}$ originating from a dynamic charge
disproportion;\cite{Brouet00} and b) the existence of a non-magnetic
insulating phase in (NH$_3$)$_2$NaK$_2$C$_{60}$,\cite{Riccounpub}
interpreted in terms of a possible (dynamic) charge disproportion.
Polaronic instabilities are in general expected at high values of the
electron-phonon coupling constant ($\lambda_c\sim
1.5$--2).\cite{Paci02}  In the case of fullerides, though, the coupling 
values suggested by different experimental techniques (even though still 
under debate\cite{Gunnarsson01}) settle in a $0.5<\lambda_{c}<1.2$ range.
The closeness of the upper limit of this range to that of the previously 
mentioned polaronic superconductivity boundary could suggest a possible
role played by polarons in these systems.

\section{Conclusions}
\label{sec:conclusions}
In this work we have shown that the magnetic properties of the 
fullerene based superconductor \nhc\ are quite different from those 
of other fullerides, namely it displays a very small $H_{c1}$ associated 
with very large $H_{c2}$ and $\lambda$ values.
These results cannot be explained within the framework of the 
Ginzburg-Landau (GL) theory. It is suggested that the system considered 
here represents a border-line case, in which the nature of 
superconducting coupling begins to switch from a phonon-mediated
to a polaronic character.

Indeed, electron-phonon interaction could be sufficiently large in general to 
make it possible for some systems, like (NH$_3$)$_x$NaK$_2$C$_{60}$, to 
exhibit an enhancement of the coupling which, in turn, favours the 
development of polaronic instabilities. In the present situation, specific
quantitative predictions for measurable superconducting quantities are
needed in order to understand the experimental results and clarify the
essential nature of superconductivity in fullerides.

\begin{acknowledgments}
We thank Prof.\ E.\ H.\ Brandt who kindly provided us with the simulation 
code for muon spin depolarization calculations and Dr G.\ Salviati with 
Dr N.\ Armani who performed the SEM granulometry measurements.\\
We acknowledge the financial support by the EU Improvement of Human 
Potential (IHP) programme, which sponsored part of our $\mu$SR experiments 
at ISIS facility (UK).  
\end{acknowledgments}

\bibliography{ricco}

\end{document}